# Memristive response and neuromorphic functionality of polycrystalline ferroelectric Ca:HfO$_2$-based devices

C. Ferreyra,[1] M. Badillo,[2, 3] M. J. Sánchez,[4, 5] M. Acuautla,[3] B. Noheda,[6, 7] and D. Rubi*[1]

[1)] *Instituto de Nanociencia y Nanotecnología (INN), CONICET-CNEA, nodo Constituyentes (1650), San Martín, Argentina*

[2)] *Department of Physics, Politecnico di Milano, Piazza Leonardo da Vinci 32, 20133, Milano, Italy*

[3)] *Engineering and Technology Institute Gronigen (ENTEG), University of Groningen, Nijenborgh 4, 9747AG Groningen, The Netherlands*

[4)] *Instituto de Nanociencia y Nanotecnología (INN), CONICET-CNEA, nodo Bariloche (8400), San Carlos de Bariloche, Argentina*

[5)] *Instituto Balseiro, UNCuyo (8400), San Carlos de Bariloche, Argentina*

[6)] *CogniGron - Groningen Cognitive Systems and Materials Center, University of Groningen, Nijenborgh 4, 9747AG Groningen, The Netherlands*

[7)] *Zernike Institute for Advanced Materials, University of Groningen, Nijenborgh 4, 9747AG Groningen, The Netherlands*

(*Electronic mail: diego.rubi@gmail.com)

Memristors are considered key building blocks for the development of neuromorphic computing hardware. For ferroelectric memristors with a capacitor-like structure, the polarization direction modulates the height of the Schottky barriers -present at ferroelectric/metal interfaces- that control the device resistance. Here, we unveil the coexistence of multiple memristive mechanisms in Pt/Ca:HfO$_2$/Pt devices fabricated on silicon by a simple and effective low-toxicity chemical solution method. Depending on the fabrication conditions, either dielectric or ferroelectric devices are obtained, each one presenting a distinct memristive response. The devices are forming-free and can sustain ferroelectric switching and memristive behavior simultaneously. Aided by numerical simulations, we describe this behavior as a competition of different mechanisms, including the effect of the ferroelectric polarization on Schottky interfaces and oxygen vacancy electromigration. Finally, we propose a simple learning algorithm for time-series recognition, designed to take advantage of the resistance relaxations present in the case of the ferroelectric devices.

## I. INTRODUCTION

Ferroelectric materials are expected to play a key role in the development of novel nanoelectronic devices, including ferroelectric field-effect transistors (FeFETs)[1], ferroelectric random access memories (FeRAMs)[2] and memristive nanodevices for in-memory or neuromorphic computing[3]. For the latter, analog resistance changes of metal/ferroelectric/metal structures, driven by polarization switching by an external electric field, are needed, mimicking the adaptive synaptic weights of biological synapses.

Ferroelectric memristive effects have been reported for perovskite materials such as BaTiO$_3$ or PZT, either for ultrathin oxides layers where the direction of the polarization controls the tunneling currents and, therefore, the device resistance (the so-called Ferroelectric Tunnel Junctions, or FTJ)[4–7] or for thicker ferroelectric layers where the energy barrier height at the interfaces with the metallic electrodes can be also tuned by the polarization direction[8–10]. Oxygen vacancy electromigration has also been shown to contribute to the memristive effect on these devices[11–13], giving rise to a complex scenario with multiple, entangled, physical mechanisms. Reports on neuromorphic properties, such as long-term potentiation/depression and spike-timing-dependent plasticity, can already be found in the literature for these systems[14–16]. Ferroelectric memristive devices based on perovskites present, however, several drawbacks that hamper their transfer to the electronics market, such as their lack of compatibility with standard Complementary Metal Oxide Semiconductor (CMOS) processes and the presence of depolarizing effects at ultra-low thickness[17], which prevents device miniaturization down to the nanoscale.

The discovery of ferroelectric HfO$_2$ (hafnia)-based oxides in 2011[18] triggered a great deal of research in the community working on oxide ferroelectrics, as this compound is CMOS compatible (indeed, it has been used as a gate oxide in metal-oxide-semiconductor field-effect transistors (MOSFETs) since early 2000's[19]), it shows robust ferroelectricity for ultrathin thickness -down to 1 nm[20]- and is also lead-free and environmentally friendly. Hafnia presents a monoclinic, nonpolar, lowest free energy phase; however, several metastable phases (cubic, tetragonal, orthorhombic, rhombohedral, the last two polar) can be stabilized either in nanostructured materials or in thin film form. The stabilization of polar phases depends on a complex interplay between grain size[21,22], the presence of cationic dopants such as Si, Al, La, Zr, Ga or Y, among others (see Ref. 23 and references therein), or the presence of oxygen vacancies[24,25], together with surface and strain effects[26,27]. The electrodes' role in stabilizing the polar phases was also highlighted in the literature[18,28,29]. It has been reported that the ferroelectric phase in hafnia-related polycrystalline devices is usually the orthorhombic one[30], while the polar rhombohedral phase was initially shown for epitaxial Hf$_{0.5}$Zr$_{0.5}$O$_2$ films grown on La$_{2/3}$Sr$_{1/3}$MnO$_3$ buffered SrTiO$_3$[31,32]. Most of the reported ferroelectric memristive effects in hafnia-based devices focus on the implementation of FTJ[33–35], which require stringent fabrication protocols to achieve high-quality ultrathin epitaxial oxide films with sharp, atomically controlled, interfaces with the metal electrodes.

Non-ferroelectric memristive mechanisms in hafnia-based devices point to the formation and dissolution of conducting nanofilaments of oxygen vacancies (which locally lower the material resistivity) after the application of proper electrical stimulation[36–38]. More recently, the local core-shell structure of filaments has been disclosed by high-resolution transmission electron microscopy[39].

Filamentary memristive effects usually require, however, an electroforming step and are characterized by steep resistance transitions (especially the one from high to low resistance, named as SET), together with high cycle-to-cycle and device-to-device variations, arising from the stochastic nature of the filament formation process[40,41]. This hinders the application of these devices in neuromorphic computing devices, where analog resistive changes and low cycle-to-cycle and device-to-device variations are desired for the implementation, for example, of physical neural networks based on memristor arrays organized in cross-bars[42]. Strategies to get analog behavior with improved reliability in hafnia-based memristors, such as adding a series transistor or including in the stack an oxygen scavenging layer, have been proposed[43,44], implying a more complex fabrication process that could impact the cost and scalability of the devices. On the other hand, the development of memristive and neuromorphic capabilities in polycrystalline hafnia-based devices displaying formingfree ferroelectricity-driven resistance change mechanisms is of high interest provided the analog behavior, high speed - related to the electronic nature of ferroelectric switching-, low cost and scalability that these devices might have.

In this paper, we disclose the memristive response of polycrystalline Ca-doped hafnia thin films (Ca:HfO$_2$), fabricated through a scalable and green chemical method[45,46]. Depending on the fabrication conditions, we were able to tune the ferroelectricity of the hafnia layer, which produced distinct memristive responses, easily detectable by a simple electrical characterization protocol. Aided by numerical simulations, the effect of Schottky barrier modulation by both the ferroelectric polarization and oxygen vacancy dynamics is disclosed. Moreover, the observed scaling of the resistive states with the device area suggests area-distributed conduction paths that we locate at grain boundaries with enhanced conductivity. We finally show the how the existence of resistance relaxations can be exploited for data classification neuromorphic algorithms.

## II. METHODS

Ca:HfO$_2$ (Ca doping of 5 % at.) thin films of three layers each (with a total thickness of 54 nm) were fabricated on platinized silicon substrates by spin-coating, followed by a pyrolysis step at $T_P$ = 300 or 400 °C for 5 minutes for each layer. The films were then processed by Rapid Thermal Annealing (RTA) at 800 °C for 90 s under a 0.5 atm Ar:O$_2$ atmosphere. Further information on the fabrication protocol and conditions can be found in Refs. 45 and 46. Circular Pt top electrodes were fabricated by e-beam evaporation and shaped by standard UV lithography. The diameter of the top electrodes was in the range of 20-200 $\mu$m, with the most useful results recorded on devices with diameter $\leq$ 50 $\mu$m, which were less prone to electrical hard breakdown. Structural characterization was done by x-ray diffraction (XRD) in a Rigaku SmartLab XE x-ray diffractometer under in-plane incidence geometry. Ferroelectricity was measured with an aixACCT TF 2000 Analyzer. Capacitance-voltage and remnant resistance-voltage loops were recorded in AC mode, simultaneously, with an LCR BK894 impedance analyzer, set for measuring an RC parallel circuit. This allows circumventing issues that arise from the highly insulating nature of our hafnia-based devices, which require measuring DC currents below the sensitivity of our instruments. Electrical characterization was performed in a commercial probe station, where the bottom electrode was grounded, and electrical stimulation was applied to the top electrode.

## III. EXPERIMENTAL RESULTS

For samples pyrolyzed at 300 °C, XRD characterization - displayed in Figure S2- shows the presence of high-symmetry fluorite phase(s) peaks (we assess that due to similar interplaner distances, cubic, tetragonal, and polar orthorhombic polymorphs can be hardly distinguished from x-ray diffraction experiments) and negligible presence of monoclinic phase[46]. We also observed, from specular XRD scans, that the films are preferentially oriented along the out-of-plane (001) direction, suggesting a fiber texture. From atomic force microscopy (AFM) analysis (Figure S1), root-mean-square (RMS) roughness was estimated to be 0.75 nm, with an average grain size of 79 nm. Figures 1(a) and (b) show typical current-voltage (I-V) and polarization-voltage (P-V) curves for these samples, both in their virgin state and after applying a wake-up protocol, consisting of 10$^3$ bipolar rectangular cycles, with an amplitude of 15 V at a frequency of 1 kHz. A dielectric behavior is evidenced for the initial cycles, while ferroelectricity develops after the wake-up process, when a polarization of 7.2 $\mu$m/cm$^2$ is obtained. We notice that the wake-up process in hafnia-based systems has been associated either to oxygen vacancy redistribution between the interface(s) and the bulk oxide zone[47], as proved by phase contrast scanning transmission electron microscopy and energy dispersive spectroscopy experiments[48], or to a phase transition between non-polar and polar polymorphs, as reported from combined in-operando Raman, photoluminescence and darkfield spectroscopy[49]. For the present case, we observed by transmission electron microscopy (to be reported elsewhere) that stressed devices present a hafnia layer with lower oxygen stoichiometry than the pristine film, suggesting that the ferroelectricity observed after wake-up is likely related to the growth of the polar orthorhombic phase[45,46] assisted by the presence of oxygen vacancies (we recall that the metastable orthorhombic phase was reported to present lower free energy in the presence of oxygen vacancies[24]).

When the pyrolysis temperature of synthesis, $T_P$, is increased to 400 °C, the Ca:HfO$_2$ films change their (001)-

oriented fiber texture to polycrystalline, as shown by the (XRD) measurements displayed in Figure S2. High-symmetry phases still make most of the Ca:HfO$_2$ film, yet evidence of monoclinic phase in small amounts is also found. A roughness of 0.52 nm and an average grain size of 102 nm were estimated for this sample (Figure S1). Noteworthily, even after the wake-up process, these devices do not display ferroelectricity (see Figure 1(c)). Understanding the origin of the correlation between texture ((001)-oriented or polycrystalline), pyrolysis temperature, and ferroelectricity is beyond the scope of the current paper and will be addressed elsewhere.

Next, we describe the capacitance (C) and memristive measurements (R$_{REM}$-V). The stimulation protocol consisted of applying alternating writing and reading DC pulses, both with a superimposed small AC signal (AC frequencies, $f$, ranged between 10 kHz and 1 MHz). The time-width of the writing pulses was $\tau_W \approx 100$ ms, and their amplitudes ranged between -18 V and +18 V, with a voltage step of 300 mV. The C-V curve is constructed from the capacitance measured during the application of the writing pulses, while the remnant resistance loop is obtained from the resistances measured during the application of the reading pulses. Both the DC bias of the reading pulses and the AC amplitude were a few hundred mV.

Figure 1(d) displays C-V loops recorded on the sample pyrolyzed at 300 °C in its original pristine state (before any electrical stress was applied) for three consecutive full cycles. It is seen that after an erratic behavior comprising a fraction of the first cycle, a butterfly-like C-V loop -typical response of ferroelectrics- is stabilized, indicating that a few pulses of the capacitive/memristive measurement protocol are enough to wake up the ferroelectric behavior. This is confirmed by Figure 1(e), which displays I-V and P-V curves recorded afterward, showing an evident ferroelectric behavior with a polarization of 3.2 $\mu$m/cm$^2$, that is 45 % of the polarization value obtained after waking-up, in Figure 1(b). The reason why the wake-up effect is induced by a few pulses of the memristive characterization protocol relies on the longer widths of the applied pulses (100 ms per pulse) in comparison to the pulses applied during the ferroelectric wake-up (1 ms per pulse). A rough comparison between both electrical protocols (memristive and ferroelectric wake-up, respectively) shows that during one full cycle of memristive cycling, the system is stressed at an average voltage of +9 V for 12 s; on the other hand, during the ferroelectric wake-up process the system is stressed at +15 V for only 1 s, that is a time-scale one order of magnitude lower than in the previous case. This suggests that the memristive protocol is strong enough to develop ferroelectricity and likely produce some fatigue, which explains the lower polarization value observed in devices after memristive characterization vs. only woken-up devices.

The R$_{REM}$-V loop of Figure 1(d) shows the presence of a pseudo-symmetric loop with a memristive butterfly that resembles an inverted C-V loop or what has also been referred to as "table-with-legs shape"[50]. The loop can be explained as arising from two memristive interfaces (likely the Ca:HfO$_2$/Pt ones, expected to be of Schottky-type) behaving in a complementary way (that is, when one switches from high to low resistance, the other switches inversely)[50,51]. We notice that the minima in the R$_{REM}$-V loop match the peaks of the C-V curve, corresponding to the coercive fields. This indicates that ferroelectric switching plays a key role in the memristive effect. We also stress that no electroforming process -usually implying a steep high-to-low resistance change- is necessary to observe this memristive effect.

The C-V and R$_{REM}$-V loops corresponding to the (non-ferroelectric) device pyrolyzed at 400 °C are displayed in Figure 1(f). The C-V loop shows roughly constant capacitance values around 21 pF, reflecting the absence of ferroelectricity. The R$_{REM}$-V loop resembles the one obtained for the ferroelectric device but with some non-trivial differences. The main difference between the remnant loops displayed in Figure 1(d) and (f) relies on the top parts of the "tables", highlighted with yellow stripped bands: while the ferroelectric sample shows distinct negative (positive) slopes for positive (negative) stimulation -resembling the behavior of BaTiO$_3$ and PZT ferroelectric perovskites[12]-, the one pyrolyzed at 400 °C shows reduced -closer to 0, in average- slopes. The existence of ferroelectricity gives rise to a linear voltage dependence of the resistance when the voltage is reduced from the maximum amplitude down to zero. We will discuss this in more detail later.

Further information about the memristive mechanism can be obtained from the evolution of the maximum (R$_{MAX}$) and minimum resistance (R$_{MIN}$) states, extracted from the R$_{REM}$-V loops, as a function of the device area. Figures 2(a) and (b) show these data for samples pyrolyzed at 300 and 400 °C, respectively, and for AC measuring frequencies in the range 10 kHz-1 MHz. We notice that the frequency-dependent character of R$_{REM}$ is related to the AC-mode in which we measure the resistance: R$_{REM}$ and C are measured with an LCR-meter assuming a parallel RC circuit, which is the equivalent representation of a usually more complex primary circuit linked to the different device zones (i.e., each metal/insulator interface is generally modeled by a capacitor in parallel with a resistor). In this way, the extracted R$_{REM}$ is frequency-dependent as it depends on capacitive reactances. In all cases, it is observed that R$_{MAX}$ and R$_{MIN}$ scale with the inverse of the device area, indicating that electrical transport is distributed along the entire device area. This allows discarding a memristive effect of filamentary nature, and it is consistent with the absence of a forming step, discussed above. This is surprising because filament formation is expected in materials with large bandgap, and it is, indeed, usually reported for hafnia-based devices[37,38]. It is also seen that the R$_{MAX}$/R$_{MIN}$ ratio is consistently in the 2-3 range, which is smaller than the usually large ratios observed in filamentary-related memristive effects[52]. For completeness, Figures 2(c) and (d) display the evolution of the low voltage capacitance (extracted from the C-V curves) as a function of the device area and measuring AC frequency. For devices pyrolyzed at 300 and 400 °C, a linear scaling of the capacitance with the area is observed, in agreement with the expected behavior of a standard capacitor.

It could be argued that the observed behavior is consistent with an interface-related memristive mechanism, where Schottky barriers present at ferroelectric/metal interfaces are modulated

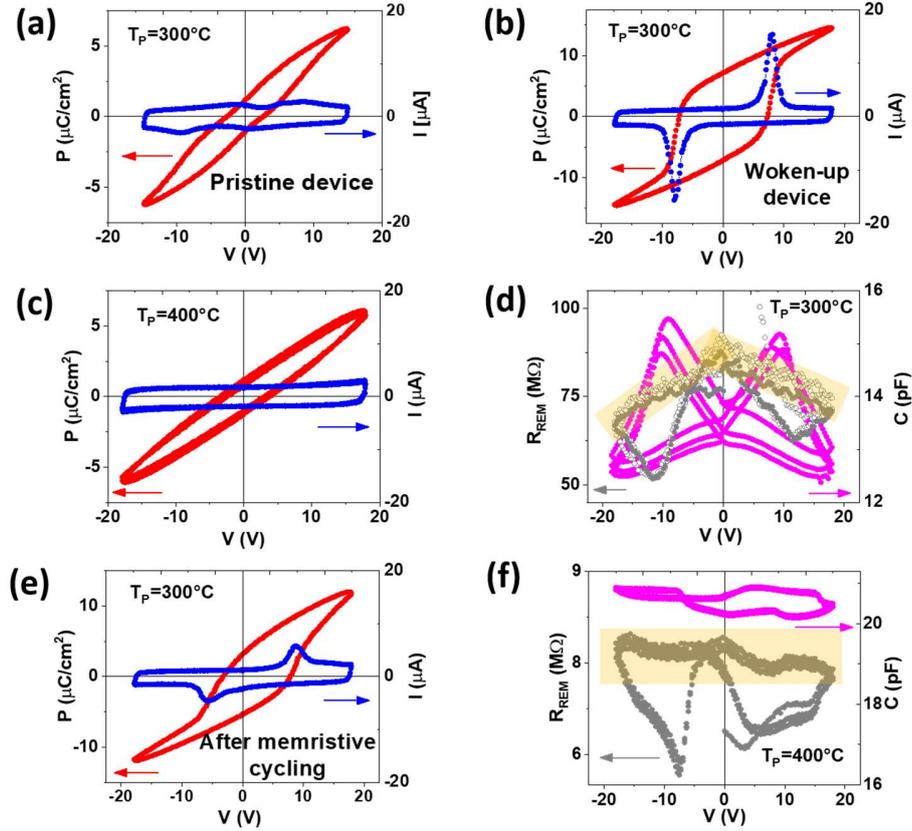

FIG. 1. (a) Current-voltage (blue symbols) and polarization-voltage curves (red symbols) recorded on an as-prepared Pt/Ca:HfO$_2$ device (diameter 35 $\mu$m) pyrolyzed at 300 °C. The measurement frequency was 1 kHz; (b) Current-voltage (blue symbols) and polarization-voltage curves (red symbols) were recorded on the same device after the application of the wake-up protocol. Ferroelectricity develops after wake-up (10$^3$ cycles). The measurement frequency was 1 kHz; (c) Current-voltage (blue symbols) and polarization-voltage curves (red symbols) recorded on an as-prepared Pt/Ca:HfO$_2$ device (electrode diameter 50 $\mu$m) pyrolyzed at 400 °C. The measurement frequency was 1 kHz. No changes in the dielectric response are observed after the application of the wake-up process; (d) Capacitance-voltage (magenta symbols) and remnant resistance-voltage (gray symbols) curves recorded on an as-prepared Pt/Ca:HfO$_2$ device (with electrode diameter of 35 $\mu$m) pyrolyzed at 300 °C. The AC frequency of the stimulation was 10 kHz; (e) Current-voltage (blue symbols) and polarization-voltage curves (red symbols) recorded on the same device, after the characterization displayed in panel (d), where the development of ferroelectricity is also observed. The measurement frequency was 1kHz; (f) Capacitance-voltage (purple symbols) and remnant resistance-voltage (gray symbols), recorded on an as-prepared Pt/Ca:HfO$_2$ device (electrode diameter 50 $\mu$m) pyrolyzed at 400 °C. The AC frequency of the stimulation was 100 kHz

by both the direction of the ferroelectric polarization and oxygen vacancy electromigration[12]. The highly insulating nature of hafnia (with an electronic bandgap of $\approx$ 6 eV) suggests, however, that electronic transport might not take place homogeneously along the complete device area but that it probably occurs through grain boundaries (we recall the fiber texture and polycrystalline nature of samples pyrolyzed at 300 °C and 400 °C, respectively), which are usually more conducting than the bulk material[53]. In addition, it has been shown that grain boundaries behave as easy migration paths for oxygen vacancy hopping[54–57]. Therefore, for the case of the sample pyrolyzed at 300 °C, we postulate that the observed (forming-free) memristive mechanism relies on oxygen vacancy dynamics through grain boundaries, entangled with the effect of ferroelectric polarization on the interface barrier heights. We can model this as follows.

## IV. MEMRISTIVE BEHAVIOR MODELLING

We describe the memristive response of both samples (the ferroelectric and the non-ferroelectric one) by using the Voltage Enhanced Oxygen Drift (VEOD)[50] model adapted to ferroelectric materials, developed in Ref. 12 to model BaTiO$_3$ and PZT-based devices. The model assumes that the device resistance is dominated by two complementary (left (L) and right (R)) Schottky-like interfaces, presenting high energy barriers which are the consequence of both the n-type character of hafnia-based oxides and the large work function of Pt, $\Phi \approx$ 5.6 eV. The barrier heights are modulated both by the direction of the polarization (if it points to (from) the interface, it lowers (increases) the barrier given by the difference between the metal work function and the electron affinity of the insulating oxide) and by the local concentration of oxygen vacan-cies. In addition, the central zone C represents the "bulk" of the memristive device. The inset of Figure 3(a) shows a sketch of the device representation used in the simulations, where L, C, and R zones are indicated.

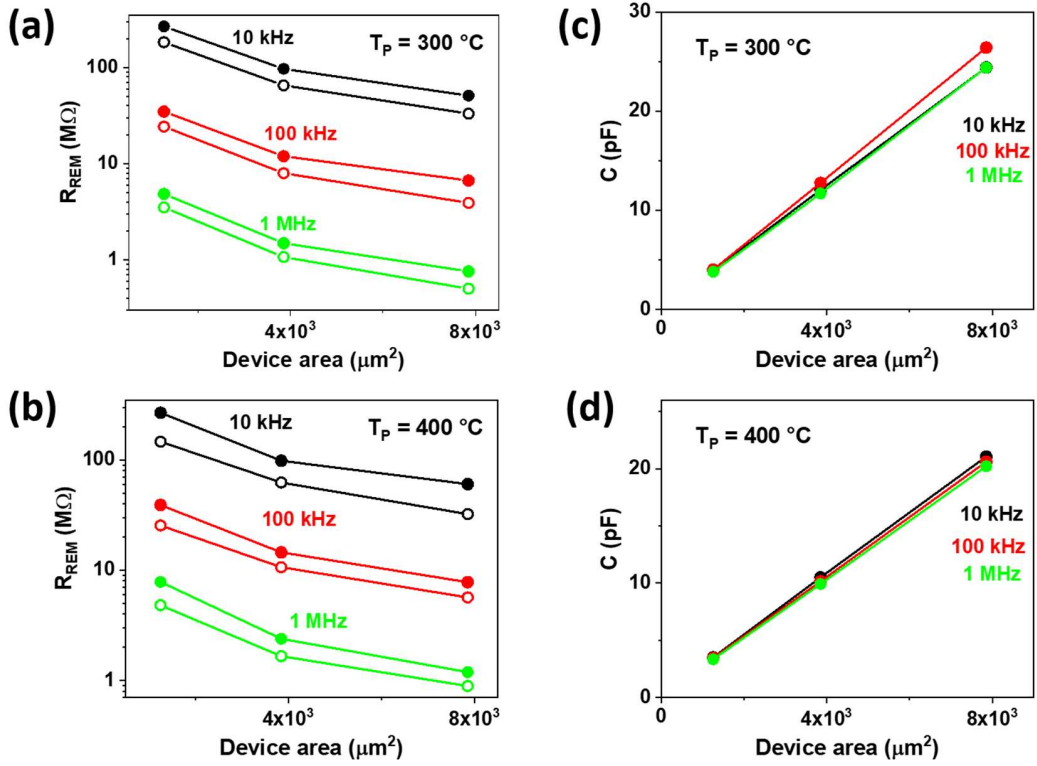

FIG. 2. Evolution of the maximum ($R_{MAX}$, full circles) and minimum ($R_{MAX}$, open circles) remnant resistance ($R_{REM}$) as a function of the device area and for different AC measuring frequencies (see main text for further details about the measurement protocol), for devices pyrolyzed at (a) 300 °C (ferroelectric devices) and (b) 400 °C (non-ferroelectric devices). Panels (c) and (d) display the corresponding capacitances as a function of the frequency and the device area.

The model assumes a 1D network of N nanodomains - formed by $N_L$ domains at the L interface, $N_R$ domains at the R interface, and $N_C$ central domains-, each one presenting a different local concentration of oxygen vacancies ($\delta_i$) that controls the local domain resistivity ($\rho_i$). Vacancies can migrate to/from neighbor domains driven by the local voltage drop at the domain, which is determined by the combination of the externally applied stimulation and the depolarizing field arising from the incomplete screening of the ferroelectric bound charges.[12].

Assuming that the two-point resistance is given by

$$R^T = R^{eff} A^{FE}(P) M^{OV},  \qquad (1)$$

where $R^{eff}$ is a scaling factor and

$$A_{FE} \equiv \begin{cases} \frac{1}{\exp(-\tilde{\gamma}_R P)} & if \ P \geq 0 \\ \frac{1}{\exp(\tilde{\gamma}_L P)} & if \ P < 0 \end{cases} \qquad (2)$$

accounts for the ferroelectric modulation of oxide-metal Schottky interfaces.

In the simulations, we take P as the polarization obtained from the experimental P-V loop, being $\tilde{\gamma}_L$ and $\tilde{\gamma}_R$ parameters that account for the barrier height correction at L and R interfaces, respectively, due to the polarization[12]. In addition, the prefactor

$$M^{OV} = N - \sum_{i=1}^{NL} A_L \delta_i - \sum_{i=NL+1}^{NL+NC} A_C \delta_i - \sum_{i=N-NR+1}^{N} A_R \delta_i, \qquad (3)$$

accounts for the oxygen vacancy dynamics, where $A_i$ is a proportionality factor that correlates the local density of oxygen vacancies $\delta_i$ with the local resistivity $\rho_i$, following the relation $\rho_i \propto (1 - A_i \delta_i)$ usually considered for devices in which the local resistivity decreases with the increase in the oxygen vacancies content (and satisfying $A_i \delta_i \ll 1$[58]). In the present case, we consider $A_i = A_L, A_C, A_R$, inside each region L, R and C, respectively.

Assuming an initial oxygen vacancy profile, the model allows calculating the vacancy hopping transfer rate $p_{i\,j}$ between adjacent domains as

$$p_{ij} = \delta_i (1 - \delta_j) \exp(-V_0 + \Delta V_i) \qquad (4)$$

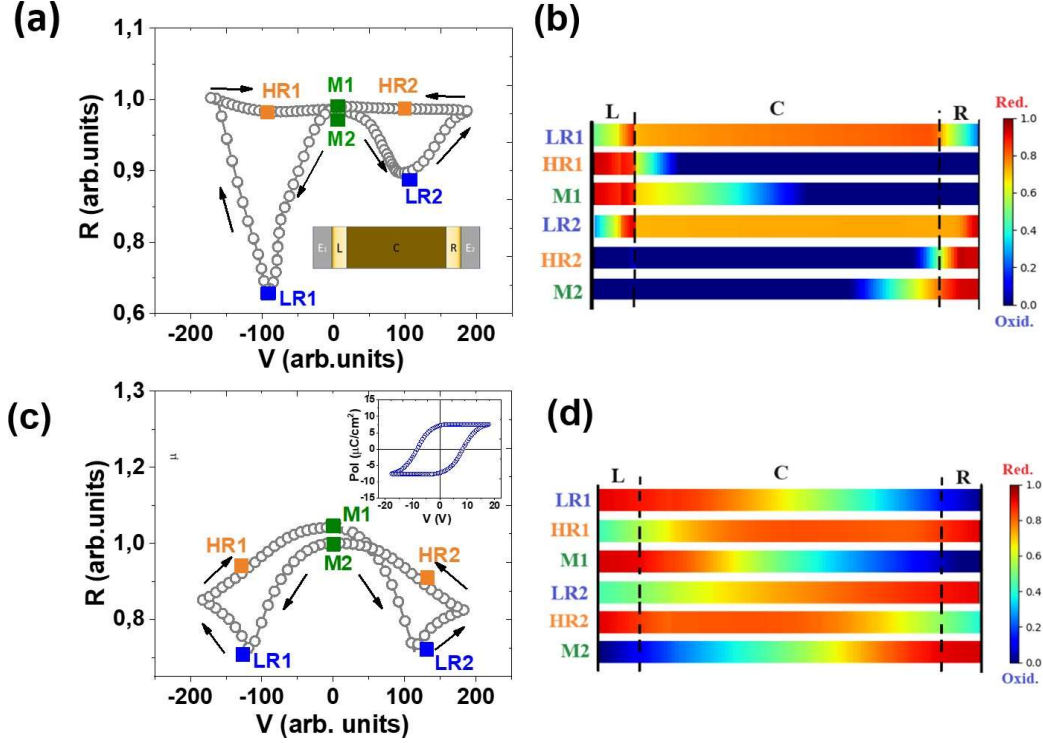

FIG. 3. Simulated remnant resistance loops for the samples pyrolyzed at (a) 400 °C (non-ferroelectric) and (c) 300 °C (ferroelectric). The inset of panel (a) displays a sketch of the device assumed for the simulation. The inset of panel (c) shows the polarization-voltage loop of the ferroelectric sample integrated from the capacitance-voltage response. The evolution of the curves is M1→LR2→HR2→M2→LR1→HR1→M1. Panels (b) and (d) display oxygen vacancy profiles (color maps) corresponding to the resistance states marked in panels (a) and (c).

with $V_0$ being the activation energy for vacancy hopping between neighbor domains, that we take without loss of generality as a constant along the device, and $\Delta V_i$ is the local voltage drop expressed as

$$\Delta V_i = \Delta V_i^W - \xi P,$$ (5)

where $\Delta V^W_i$ is due to the external write voltage ($V^W$), and the second term takes into account the voltage drop due to the depolarizing field[59] $E_{DP} = -\xi P$, with the parameter $\xi$ tuning its strength.

After calculating the vacancy hopping rates for each simulation step, the device resistance is updated. This procedure is repeated iteratively for an external voltage ramp with symmetric voltage excursions between $V_{MAX} = +1800$ arb. units and $V_{MIN} = -1800$ arb. units (20 arb. units correspond to $\approx$ 1V), respectively, until a full R-$V^W$ loop is obtained. The time-width of the write pulses was set in 10 arb. units, which corresponds to $\approx$ 100 ms. We refer to Ref. 12 for further details on the model and its implementation.

Figure 3(a) shows the simulated remnant resistance loop corresponding to the device pyrolyzed at 400 °C (non-ferroelectric). For this sample, $A^{FE} = 1$ was assumed. The numerical values given to the model parameters are shown in the Supplementary Information, Table S1. The simulated remnant resistance loop displays a "table-with-legs" shape that comprises the main features observed experimentally (see Figure 1(f)); in particular, the top part of the "table" is flat, reflecting the time stability of the high resistance state. The asymmetry between positive and negative stimulation resembles the experimental loop (recall Figure 1(f)). It is related to the fact that, despite the nominal symmetry of both ferroelectric/metal interfaces, they present different thermal histories -the bottom interface (R in the sketch of Figure 3(a)) was annealed at high temperature while the top one (L in the sketch), fabricated ex-situ, was not-. The asymmetry between both interfaces is taken into account by our model by giving slightly different values to the $A_L$ and $A_R$ parameters (see Table S1). We also notice that $A_L$, $A_R > A_C$, which is necessary to warrant a higher resistivity in the zones close to the interfaces (L and R) compared to the central (C) zone, consistently with their Schottky nature.

Figure 3(b) shows oxygen vacancy density heat maps corresponding to the resistance states indicated in Figure 3(a). We stress that, in this case, the memristive behavior is solely related to the internal vacancy electromigration between L and R interfaces, as it was shown in standard memristive systems such as manganites[50] or TiO$_x$ and TaO$_x$-based[12,60] devices.

Figure 3(c) displays the simulated remnant resistance loop corresponding to the ferroelectric device. In this case, the $A^{FE}$ factor was considered as in Eq.(2). The P-V loop, used as an input for the calculation of $A^{FE}$, is displayed as an inset and was obtained after time integration of the experimental C- V response. The simulated remnant resistance loop presents the following features, in line with the experimental loop displayed in Figure 1(c): i) the "legs" of the "table" match the coercive fields, reflecting that polarization reversal plays a fundamental role in the memristive effect; ii) the top part of the "table" displays non-zero slopes, which are an indication of resistance relaxations (volatile effect) in the absence of external stimulation, as we have also shown for Pt/PZT/Pt ferroelectric devices[12]. In our model, these relaxations are attributed to oxygen vacancy electromigration driven by the depolarizing field $E_{DP}$ (recall Eq.(5)).

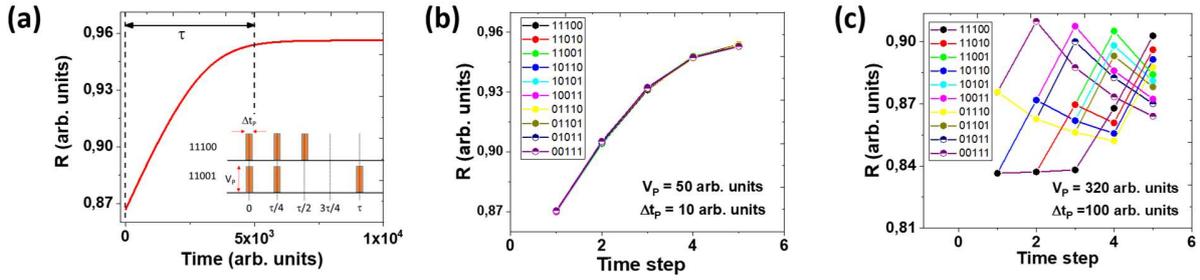

FIG. 4. (a) Simulation of the remnant resistance evolution with time under zero external bias, starting from the HR2 state displayed in Figure 3(c). The observed resistance relaxation is related to oxygen vacancy migration in the absence of external stimulation. The typical time-length of the relaxation is indicated as $\tau$; (b) Simulation of the remnant resistance vectors ($R_1$, $R_2$, $R_3$, $R_4$, $R_5$), obtained after evaluating the remnant resistance at $t_1 = 0$, $t_2 = \tau/4$, $t_3 = \tau/2$, $t_4 = 3\tau/4$, and $t_5 = \tau$ upon the application of 10 different pulse time series (see the inset of panel (a), where two series are displayed as an example). The pulse amplitude and time-width were $V_P = 50$ arb. units and $\Delta t_P = 10$ arb. units, respectively. See the main text for further details; (c) Simulation of the remnant resistance vectors, under the same pulse scheme of panel (b), now with $V_P = 320$ arb. units and $\Delta t_P = 100$ arb. units. In this case, all remnant resistance vectors are fully distinguishable from each other.

We notice that the existence of a robust $E_{DP}$ in hafnia-related oxides remains controversial. It can be argued that the relatively high polarization found in ultra-thin hafnia films might originate from a low $E_{DP}$, possibly screened by point defects such as oxygen vacancies. However, other reports suggest the existence of a strong $E_{DP}$ in hafnia-based devices[61]. Our results are more aligned with the latter scenario, as it assumes the presence of a $E_{DP}$ strong enough to participate in oxygen vacancy dynamics but lower than the coercive field, avoiding in this way depolarization effects.

The numerical values used in modeling the remnant resistance loop of Figure 3(c) are listed in Table S1. Figure 3(d) displays oxygen vacancy density heat maps corresponding to the resistance states indicated in the remnant resistance loop of Figure 3(c). The comparison with the non-ferroelectric device (Fig. 3(b)) suggests softer oxygen vacancy profiles for the ferroelectric case (milder color gradients). This can be explained in terms of the competing effects on oxygen vacancy dynamics due to both the external electric field and $E_{DP}$; for instance, a positive external voltage applied to the electrode next to the L interface intends to move (positively charged) vacancies towards the R interface; however, as the sample is polarized pointing to the R interface, the $E_{DP}$ points to the opposite direction and, therefore, tries to pump vacancies towards the L interface, competing with the effect of the external stimulation.

## V. NEUROMORPHIC FUNCTIONALITY: TIME SERIES CLASSIFICATION

The presence of resistance relaxations is of high interest for neuromorphic devices. Figure 4(a) simulates the evolution of the resistance under zero external voltage for a ferroelectric device initially set in a high resistance state (close to the HR2 state of Fig. 3(c)). It can be seen that the resistance progressively increases as the system evolves. The relaxation is similar to what was experimentally found in PZT-based devices[12], and it is related to oxygen vacancy dynamics driven by $E_{DP}$, presenting a characteristic time $\tau \approx 5x10^3$ arb. units, which corresponds to $\approx$ 50 s. Next, we simulate a possible application of this property for time-series classification.

Starting in all cases from the resistance state HR2, we simulated the evolution of the remanent resistance of the ferroelectric device upon the application of series of 3 pulses (each one with amplitude $V_P$ and time-width $\Delta t_P$) distributed in 5- time steps ($t_1 = 0$, $t_2 = \tau/4$, $t_3 = \tau/2$, $t_4 = 3\tau/4$, $t_5 = \tau$), where $\tau$ is defined as the time-length of the resistance relaxation, as indicated in Figure 4(a). Within this scheme, there are ten possible different sequences of pulses, 2 of which are displayed as examples in the inset of Figure 4(a) ("1" and "0" indicate if the pulse has been applied or not in each time step). The simulation determines the remnant resistance immediately after each time-step (either if a pulse was applied or not), and a remnant resistance vector ($R_1$, $R_2$, $R_3$, $R_4$, $R_5$), where $R_i$ corresponds to the remnant resistance at $t^*$, is constructed. We notice that in the case of a purely non-volatile memristor, the resistance vector, for a given value of $V_P$ and $\Delta t_P$, should be the same for all the pulsing sequences, as the resistance would evolve according to the potentiation-depression curves of the system, independently of the timing between pulses. The situation is different for a memristive system displaying relaxations, as is the case of our ferroelectric device. In this case, there will be a trade-off between resistance changes induced by the external pulses and the relaxation between pulses, leading to different remnant resistance vectors depending on the timing of the applied pulses. Figure 4(b) shows the resistance vectors for the different pulse series, for the case of pulses with $V_P = 50$ arb. units and $\Delta t_P = 10$ arb. units. In this case, it is seen that the pulses are not strong enough to induce resistance changes, and the resistance evolution follows the free relaxations displayed in Figure 4(a). When the strength of the pulses is enlarged to $V_P = 320$ arb. units and $\Delta t_P = 100$ arb. units the mentioned trade-off occurs, and the resistance vectors corresponding to the different pulse time series are fully distinguishable from each other, as a close inspection of Figure 4(c) evidences. This behavior relies on the non-linear nature of the resistance relaxations. It has been proposed that this effect can be exploited for neuromorphic algorithms, where the resistance vectors feed a fully connected readout layer with trainable synaptic weights, in order to complete the data set classification process[62]. The results presented above show that our hafnia-based memristors would be appropriate for the physical implementation of time-series classification algorithms, with potentially significant energy saving due to the substantially reduced amount of trainable parameters compared to a standard neural network.

## VI. CONCLUSIONS

In summary, we have shown that the ferroelectricity of Ca-doped hafnia-based devices, fabricated by a simple and effective low-toxicity chemical solution method, can be tailored by subtly modifying fabrication parameters such as the pyrolysis temperature during synthesis. Unlike other reports on hafnia-based memristors, our devices are forming-free and show area-dependent memristive behavior, which is inconsistent with filament formation. In addition, we obtain ferroelectric and non-ferroelectric devices, with distinct memristive responses, depending on the history of the processing conditions (i.e., pyrolysis temperature). The devices display both ferroelectric and memristive behavior in parallel, unlike previous reports on epitaxial samples, which showed that the two effects occur independently[52]. To explain our results, we propose a physical scenario where both the ferroelectric polarization direction and the oxygen vacancy dynamics modulate the Schottky barrier heights. In absence of external stimulation, resistance relaxations occur and we show that this effect could be exploited for the development of neuromorphic hardware.

### Acknowledgments

We acknowledge support from ANPCyT (PICT2019-02781, PICT2020A-00415) and EU-H2020-RISE project MELON (Grant No. 872631). This work was also supported by The National Council for the Humanities, Science, and Technology (CONAHCYT, México) under a Postdoc grant to Miguel Badillo (CVU 356403). We are also grateful to NanoLab and Polifab facilities and their staff at the University of Groningen and Politecnico di Milano. Financial support by the Groningen Cognitive Systems and Materials Centre (CogniGron) and the Ubbo Emmius Foundation of the University of Groningen is gratefully acknowledged.

### Supplementary Material

See Supplementary Material for atomic force microscopy and x-ray diffraction characterization and for further details about the model that simulates the memristive properties.

### Conflict of Interest

The authors have no conflicts to disclose.

### Data availability

The data supporting this study's findings are available from the corresponding author upon reasonable request.